\documentclass[twocolumn,superscriptaddress]{revtex4-1}
\usepackage{lmodern}
\usepackage{lmodern}
\usepackage[utf8x]{inputenc}
\usepackage{color}
\usepackage{verbatim}
\usepackage{amsmath}
\usepackage{graphicx}

\makeatletter
\usepackage{float}

\usepackage{amsfonts}
\usepackage{url}
\usepackage{epstopdf}
\usepackage{placeins}
\usepackage{xcolor}

\makeatother

\begin{document}
\title{Dynamically Generated Synthetic Electric Fields for Photons}
\author{Petr Zapletal}
\affiliation{Cavendish Laboratory, University of Cambridge, Cambridge CB3 0HE,
United Kingdom}
\affiliation{Max Planck Institute for the Science of Light, Staudtstraße 2, 91058
Erlangen, Germany}
\author{Stefan Walter}
\affiliation{Max Planck Institute for the Science of Light, Staudtstraße 2, 91058
Erlangen, Germany}
\affiliation{Institute for Theoretical Physics, University Erlangen-Nürnberg, Staudtstraße
7, 91058 Erlangen, Germany}
\author{Florian Marquardt}
\affiliation{Max Planck Institute for the Science of Light, Staudtstraße 2, 91058
Erlangen, Germany}
\affiliation{Institute for Theoretical Physics, University Erlangen-Nürnberg, Staudtstraße
7, 91058 Erlangen, Germany}
\begin{abstract}
Static synthetic magnetic fields give rise to phenomena including
the Lorentz force and the quantum Hall effect even for neutral particles,
and they have by now been implemented in a variety of physical systems.
Moving towards fully dynamical synthetic gauge fields allows, in addition,
for backaction of the particles' motion onto the field. If this results
in a time-dependent vector potential, conventional electromagnetism
predicts the generation of an electric field. Here, we show how synthetic
electric fields for photons arise self-consistently due to the nonlinear
dynamics in a driven system. Our analysis is based on optomechanical
arrays, where dynamical gauge fields arise naturally from phonon-assisted
photon tunneling. We study open, one-dimensional arrays, where synthetic
magnetic fields are absent. However, we show that synthetic electric
fields can be generated dynamically. The generation of these fields
depends on the direction of photon propagation, leading to a novel
mechanism for a photon diode, inducing nonlinear unidirectional transport
via dynamical synthetic gauge fields. 
\end{abstract}
\maketitle
The field of cavity optomechanics, addressing the interaction between
light and sound, has made rapid strides in recent years \citep{Aspelmeyer2014}.
Experiments have shown ground state cooling \citep{chan2011,teufel2011},
measurements of motion with record sensitivity \citep{wilson2015},
efficient conversion between microwave and optical photons \citep{Andrews2014},
dynamics of vibrations near exceptional points \citep{xu2016}, and
the control of single phonons \citep{hong2017}, to name but a few
achievements. 

Due to the optomechanical interaction, mechanical vibrations can change
light frequency. During this process, the mechanical oscillation phase
is imparted onto light field. This provides a natural means to generate
synthetic magnetic fields for photons, as was first suggested in Refs.~\citep{hafezi2012,schmidt2015}.
Together with reservoir engineering \citep{metelmann2015}, these
ideas form the theoretical basis underlying a recent series of pioneering
experiments on optomechanical nonreciprocity \citep{kim2015,wang2015,ruesink2016,fang2017,bernier2017,barzanjeh2017}.
While those still operate in few-mode setups, future extensions to
optomechanical arrays \citep{chang2011,heinrich2011,xuereb2012,ludwig2013}
will enable studying photon transport on a lattice in presence of
an arbitrary tunable synthetic magnetic field \citep{schmidt2015}.
A similar optomechanical design underlies the first proposal for engineered
topological phonon transport \citep{peano2015}. These developments
tie into the wide field of synthetic magnetic fields and topologically
protected nonreciprocal transport, first envisaged and implemented
for cold atoms \citep{jaksch2003,lin2009,aidelsburger2011,aidelsburger2013}
and then for photons \citep{haldane2008,wang2009,hafezi2011,fang2012,rechtsman2013,hafezi2013,mittal2014},
phonons \citep{peano2015,nash2015,susstrunk2015,wang2015a,brendel2017,brendel2018,seif2018},
and other platforms \citep{goldman2014,hartmann2016}.

In these works, gauge fields are fixed by external parameters, e.g.,
the phases of external driving beams. It was understood only recently
that optomechanics provides a very natural platform for creating \emph{dynamical}
classical gauge fields \citep{walter2016}: if the mechanical resonator
is not periodically modulated by external driving but rather undergoes
limit-cycle oscillations, the phase of those oscillations becomes
a dynamical gauge field. This field is a new degree of freedom that
can be influenced by photons.

The theory of \textit{classical} dynamical gauge fields is not only
important as a starting point for high-energy \textit{quantum} field
theory (e.g. Yang-Mills and Higgs theories \citep{yang1954,anderson1963,higgs1964}),
but is also an active area of research in modern condensed matter
physics (e.g. in the gauge theory of dislocations \citep{kleinert1989},
spin ice \citep{castelnovo2012} and nematic liquid crystals \citep{lammert1993,lammert1995}).
We emphasize that our main goal is different from the attempts to
build quantum simulators for existing high-energy gauge theories (suggested
theoretically for ultra-cold atoms in optical lattices \citep{banerjee2012,zohar2016},
superconducting circuits \citep{marcos2013,marcos2014}, cavity quantum
electrodynamics \citep{ballantine2017}, and trapped ions \citep{hauke2013}),
where the experimental implementation remains very challenging (see
Ref.\ \citep{martinez2016} for the first steps). Rather, our work
provides new insights for all physical systems where limit-cycle oscillators
assist transitions between linear modes, by connecting these systems
to the general mathematical framework of classical dynamical gauge
fields. This includes different kinds of limit-cycles (electrical,
mechanical, optical, and spin oscillators, pumped using electrical
feedback, optomechanical backaction, etc.), different kinds of linear
modes (microwave, mechanical, optical, magnon resonances, etc.), and
almost arbitrary nonlinear coupling, using optical or mechanical nonlinearities,
Josephson junctions, etc. For concreteness, we describe it here for
the case of optomechanics, but the mathematics and the predictions
are general and of wide experimental applicability.

If the gauge field dynamics results in a time-dependent vector potential,
conventional electromagnetism dictates that this describes an electric
field. In this work, we predict that synthetic electric fields can
arise in elementary optomechanical systems, in a dynamical way. The
scenarios in which these electric fields arise, and their physical
consequences, are qualitatively different from the more conventional
self-consistently generated magnetic fields discussed in our previous
work \citep{walter2016}. They can arise even in a linear arrangement
of coupled photon modes, where static vector potentials do not have
any effect, since they can be gauged away. This makes them a very
relevant phenomenon for present-day experimental implementations,
in setups as simple as two coupled optical modes. Moreover, the appearance
of electric fields turns out to depend on the direction of photon
propagation. In this way, we uncover a novel mechanism for nonlinear
unidirectional transport of photons (a photon diode). This works especially
well in arrays, where transport is significantly suppressed in the
blockaded direction.

Synthetic electric fields for photons have been previously analyzed
only in the context of prescribed external driving \citep{yuan2015,yuan2016},
i.e. not dynamically generated. In these cases, the nonlinear dynamics
and unidirectional transport explored in our work are absent.

\begin{figure}[t]
\centering \includegraphics[width=0.95\linewidth]{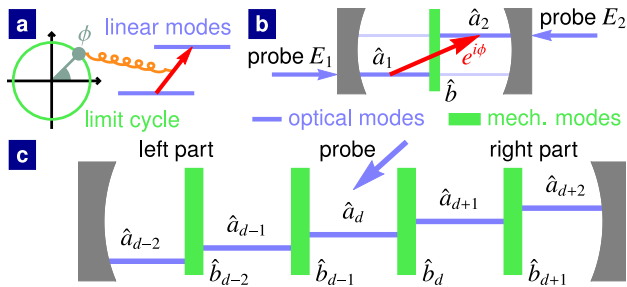} \caption{Setup exhibiting dynamically generated synthetic electric fields.
(a) In general terms: limit-cycle oscillator assisting
the transition between linear modes. (b) Optomechanical realization:
a cavity with a movable membrane (green rectangle) in the middle supporting
optical supermodes $\hat{a}_{1}$ and $\hat{a}_{2}$ (mostly localized
left and right, respectively). The mechanical mode $\hat{b}$ undergoes
limit-cycle oscillations. Photons tunneling (red arrow) from optical
mode $\hat{a}_{1}$ to $\hat{a}_{2}$ absorb a phonon from the mechanical
oscillation, thereby acquiring a phase shift set by the oscillation
phase $\phi$. Photon transport through the setup can be probed by
driving mode $\hat{a}_{1}$ or $\hat{a}_{2}$. Optical frequencies
are represented by the blue lines. (c) A one-dimensional array, with
optical modes $\hat{a}_{j}$ of increasing frequency. Mechanical modes
$\hat{b}_{j}$ assist tunneling between modes $\hat{a}_{j}$ and $\hat{a}_{j+1}$.
Some mode, $\hat{a}_{d}$, is driven by a laser (blue arrow), probing
photon transport both towards the left and right.}
\label{link} 
\end{figure}

\begin{figure*}[t]
\centering 

\includegraphics[width=0.95\linewidth]{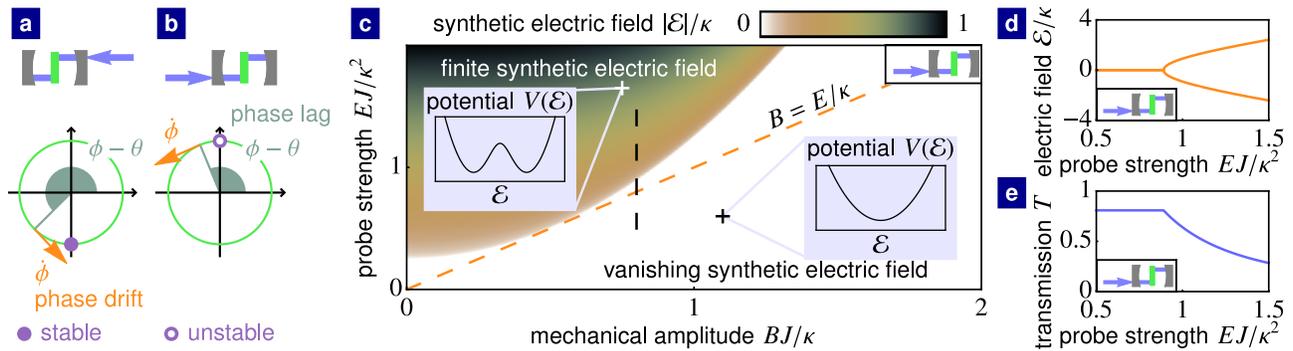} \caption{Dynamically generated synthetic electric fields in the two-site system.
(a,b) Phase evolution on the mechanical limit cycle (green orbit).
(a) When the higher optical mode is laser-driven, the system settles
into a stable fixed point, with a phase lag $\phi-\theta=-\pi/2$.
(b) When the lower optical mode is driven, the phase is continuously
repelled from an unstable fixed point $\phi-\theta=+\pi/2$, generating
a finite synthetic electric field $\mathcal{E}=\dot{\phi}\protect\neq0$
acting on the photons. (c) The phase diagram. In the white region,
$\mathcal{E}$ vanishes in the steady state. If the lower-frequency
mode, $a_{1}$, is driven, $\mathcal{E}$ bifurcates in the colored
region to finite steady-state values. Their absolute values are indicated
by the color scale. The blue insets show the effective potential $V(\mathcal{E})$
determining the steady-state value of $\mathcal{E}$. The dashed black
line denotes the cut along which $\mathcal{E}$ and the optical transmission
$T$ are plotted in (d) and (e), respectively. For the higher-frequency
mode, $a_{2}$, being driven, $\mathcal{E}$ always vanishes for any
values of the system parameters. Consequently, the transmission is
never suppressed.}
\label{phdiag} 
\end{figure*}

\emph{Dynamical gauge fields for photons}. — The optomechanical interaction
can be used to realize phonon-assisted photon tunneling, which, as
we have shown previously, offers a natural route towards classical
dynamical gauge fields for photons \citep{walter2016}. Photons hopping
between optical modes $\hat{a}_{1}$ and $\hat{a}_{2}$ absorb or
emit a phonon from a mechanical mode $\hat{b}$. A pictorial representation
of this process is shown in Fig.~\ref{link}b. Many implementations
are conceivable (photonic crystal devices, coupled toroids, and microwave
circuits \citep{Aspelmeyer2014}), but a suitable realization might
simply consist of the well-known membrane-in-the-middle setup \citep{thompson2008,sankey2010}.
The Hamiltonian is 
\begin{equation}
\hat{H}=\sum_{j=1}^{2}\nu_{j}\hat{a}_{j}^{\dagger}\hat{a}_{j}+\Omega\,\hat{b}^{\dagger}\hat{b}+J\left(\hat{b}\hat{a}_{2}^{\dagger}\hat{a}_{1}+{\rm h.c.}\right),\label{ham}
\end{equation}
where $\nu_{j}$ are the optical frequencies of modes $\hat{a}_{j}$,
$\Omega$ is the frequency of the mechanical oscillator and $J$ is
the tunneling amplitude \citep{walter2016}. In the following, we
set $\hbar=1$.
The phonon-assisted photon tunneling is selected by tuning the mechanical
frequency, $\Omega\approx|\nu_{2}-\nu_{1}|$. The Hamiltonian (\ref{ham})
is valid within the rotating-wave approximation for $\nu_{2}>\nu_{1}$
and $\Omega\gg\kappa,J$,$JB$, where $\kappa$ is the photon decay
rate and $B$ is the amplitude of the mechanical oscillations: $\left\langle \hat{b}\right\rangle =Be^{i\phi}e^{-i\Omega t}$.

During the photon tunneling process $\hat{b}\hat{a}_{2}^{\dagger}\hat{a}_{1}$,
the mechanical phase $\phi$ is imprinted on the photons, similar
to an Aharonov-Bohm (Peierls) phase. This can be used for optomechanical
generation of \emph{static} gauge fields, as proposed in Ref.~\citep{schmidt2015},
and the scheme can be readily implemented in optomechanical crystals
\citep{safavi-naeini2011,paraiso2015} or the membrane-cavity setup
\citep{thompson2008,sankey2010,wu2013}. It was experimentally realized
in Ref.\ \citep{fang2017}.

To implement \emph{dynamical} gauge fields for photons, i.e., fields
that are themselves dynamical degrees of freedom, the oscillation
phase $\phi$ (the ``gauge field'') has to evolve freely, which
is the case if the mechanical mode performs limit-cycle oscillations
\citep{walter2016}. The limit-cycle oscillations can be generated
by pumping an ancillary optical mode, situated at a different frequency,
on the blue sideband \citep{marquardt2006}. This pumping does not
impose any particular phase on the mechanical oscillator and thus
the phase is able to evolve according to its own dynamics. In this
way, the phase $\phi$ turns into a dynamical gauge field, being influenced
by photon transport and acting back on photons. The system of Eq.\ (\ref{ham})
can be used as a building block for optomechanical arrays with dynamical
gauge fields for photons, as we first proposed in Ref.\ \citep{walter2016}.
We use the equations of Ref.\ \citep{walter2016} as our starting
point, to predict the new phenomenon of synthetic electric fields
generated by nonlinear dynamics, giving rise to unidirectional photon
transport.

\emph{The basic physics behind our results}. – We start with a preview
of our results, emphasizing the physical intuition. Any oscillator
driven by a resonant force $F_{0}\cos(\Omega t-\theta)$ experiences
a drift, $\dot{\phi}\propto F_{0}\cos(\phi-\theta)$, of its phase
$\phi$. In our case, the force is the radiation pressure oscillating
at the beat note between the two optical modes, and we obtain $\dot{\phi}=-(J/B)\left|a_{1}\right|\left|a_{2}\right|\cos(\phi-\theta)$,
where $\theta$ is the phase difference between the optical modes.
If the forcing phase $\theta$ is kept constant, this results in a
stable fixed point $\phi=\theta-\pi/2$. For a limit-cycle oscillator,
that behavior is known as phase locking (injection locking) to an
external drive (see Ref.\ \citep{SupplentaryMaterial} for more details).

However, in our case an interesting self-consistency problem arises:
the phase difference $\theta=\theta_{2}-\theta_{1}$ of the two optical
modes depends on $\phi$ itself, as the phase $\phi$ is imprinted
onto the photons during the phonon-assisted photon tunneling. The
phase of the force thus follows the oscillation phase. We now discuss
qualitatively the resulting physics, which will be bolstered by detailed
analysis later. Two cases need to be distinguished, depending on which
optical mode is driven by the laser (see Fig.\ \ref{phdiag}a and
Fig.\ \ref{phdiag}b). If the \emph{higher} optical mode (labeled
'2') is driven, then we find $\theta=\phi+\pi/2$. The crucial term
$\pi/2$ comes about due to the \emph{resonant} excitation of the
lower mode via the phonon-assisted transition $2\rightarrow1$. Comparing
with the stable fixed point for $\phi$ deduced above, we conclude
that \emph{any} value of $\phi$ is now stable.

The situation drastically changes if the \emph{lower} optical mode
is driven by the laser. Then, we find $\theta=\phi-\pi/2$, where
the sign has flipped because the roles of optical modes have been
interchanged (now the higher mode is excited by the phonon sideband
of the driven lower mode). This corresponds to an unstable fixed point.
Once $\phi$ tries to move away, $\theta$ will follow, such that
$\phi$ is forever repelled. This results in a finite phase drift
$\dot{\phi}\neq0$, corresponding to an effective shift of the mechanical
frequency. Thus, the phonon-assisted tunneling process towards the
higher optical mode is no longer in resonance but detuned by $\dot{\phi}$.
This off-resonant excitation shifts the optical phase difference according
to $\theta\approx\phi-\pi/2-\dot{\phi}/(\kappa/2)$. The equation
$\dot{\phi}\propto-\cos(\phi-\theta)$ can then be fulfilled at a
certain value of $\dot{\phi}$, which has to be obtained self-consistently.
This is the qualitative origin of the nonlinear dynamics that gives
rise to what we will identify below as a synthetic electric field
$\mathcal{E}=\dot{\phi}$ acting on photons.

\emph{Dynamics and synthetic electric fields}. — Let us analyze the
dynamics of the two-site system (\ref{ham}) with the mechanical oscillator
performing limit-cycle oscillations. The optical mode $\hat{a}_{j}$
is driven by a laser of amplitude $E_{j}$ at frequency $\nu_{D,j}$,
probing photon transport through the system. The optical and mechanical
amplitudes are assumed large such that quantum noise can be neglected,
which is an excellent approximation for all existing optomechanical
experiments studying nonlinear dynamics.

Following Ref.~\citep{walter2016}, the classical equations of motion
for the optical amplitudes $a_{j}=\left\langle \hat{a}_{j}\right\rangle $
and the mechanical phase $\phi$ read 
\begin{align}
\dot{\phi} & =\Delta_{{\rm M}}-\frac{J}{B}\,{\rm Re}\left[a_{1}^{*}a_{2}e^{-i\phi}\right],\label{EOM phi}\\
\dot{a}_{1} & =i\Delta_{1}a_{1}-iE_{1}-iJBe^{-i\phi}a_{2}-\frac{\kappa}{2}a_{1},\label{EOM a1}\\
\dot{a}_{2} & =i\Delta_{2}a_{2}-iE_{2}-iJBe^{i\phi}a_{1}-\frac{\kappa}{2}a_{2},\label{EOM a2}
\end{align}
where $\Delta_{j}=\nu_{D,j}-\nu_{j}$ and $\Delta_{{\rm M}}=\nu_{D,2}-\nu_{D,1}-\Omega$
are optical and mechanical detunings, respectively (switching to suitable
rotating frames). The mechanical amplitude $B$ is considered fixed.
These equations form the starting point of our analysis.

If only one optical mode is driven, no external phase is imprinted.
The mechanical oscillator is free to pick any phase despite the interaction
with the optical modes. The phase forms a classical gauge field with
$U(1)$ symmetry. The gauge transformation 
\begin{align}
\phi & \mapsto\phi+\chi_{2}-\chi_{1},\label{gauge phi}\\
a_{j} & \mapsto a_{j}e^{i\chi_{j}},~{\rm for}~j=1,2,\label{gauge a}
\end{align}
generates a new valid solution of the dynamical equations, for any
real functions $\chi_{j}(t)$. The transformation also preserves optical
and mechanical frequencies whenever it is time-independent, i.e.,~$\chi_{j}={\rm const}$.
However, if $\chi_{j}$ are time-dependent, Eqs. (\ref{gauge phi})
and (\ref{gauge a}) have to be supplemented by a shift in frequencies:
$\Omega\mapsto\Omega+\dot{\chi}_{1}-\dot{\chi}_{2}$ and $\nu_{j}\mapsto\nu_{j}-\dot{\chi}_{j}$.
Any time-evolving phase $\phi$ can be viewed as generating a synthetic
electric field

\begin{equation}
\mathcal{E}=\dot{\phi}
\end{equation}
for photons. For example, if mode $1$ is driven, we can re-gauge
using $\chi_{1}=0,\,\chi_{2}=-\phi$, which results in a description
where the mechanical phase is static but $\nu_{2}\mapsto\nu_{2}+\mathcal{E}$.
This describes an effective optical frequency shift, which can be
interpreted as a synthetic electric field for photons in the same
way that an energy difference between electronic levels indicates
a voltage drop, i.e., a real electric field. In conventional electromagnetism,
an electric field can be represented either as a time-dependent vector
potential or a scalar potential gradient. Analogically, the synthetic
electric field $\mathcal{E}$ is described either by the time-evolution
of the mechanical phase or by an effective frequency shift. As we
will show, $\mathcal{E}$ has important consequences for photon transport. 

\emph{Dynamical phase diagram}. — Here, $\mathcal{E}$ is not prescribed
externally but it arises due to the dynamics of coupled optical and
mechanical modes. The optical modes induce the force $F$ acting on
the mechanical phase. The resulting phase evolution may generate a
field $\mathcal{E}$ which effectively modifies the optical frequency
difference and, consequently, the population of the optical modes.

The results of the dynamical analysis are shown in Fig.~\ref{phdiag}.
The results were obtained by linear stability analysis and numerical
simulations. 

We consider the fully resonant situation where physical effects are
most pronounced, as both optical driving and phonon-assisted photon
tunneling are resonant ($\Delta_{{\rm M}}=\Delta_{1}=\Delta_{2}=0$).
The system always converges to a steady state. The steady-state value
of $\mathcal{E}$ depends on two dimensionless parameters: the rescaled
limit-cycle amplitude $BJ/\kappa$ and the rescaled laser amplitude
$EJ/\kappa^{2}$. 

The dynamical analysis becomes more intuitive by ``integrating out''
optical modes. This leaves us with an effective potential $V(\mathcal{E}$),
whose minima determine steady-state values of $\mathcal{E}$ (see
Ref\ \citep{SupplentaryMaterial} for the full analytical expression):

\begin{equation}
\dot{\mathcal{E}}=-\frac{dV(\mathcal{E})}{d\mathcal{E}}=0.
\end{equation}
In the white region of the phase diagram, Fig.~\ref{phdiag}c, the
potential $V(\mathcal{E})$ has a single minimum at $\mathcal{E}=0$
(see the blue inset). For the lower-frequency mode, $a_{1}$, being
driven, this steady state becomes unstable in the colored region of
the phase diagram, where the potential $V(\mathcal{E})$ has two minima
at finite values of $\mathcal{E}$. The field $\mathcal{E}$ can develop
such a nonzero value for $B<E/\kappa$ (above the dashed orange line).
In terms of physical parameters, the occupation of the driven optical
mode has to exceed the phonon number in the limit-cycle oscillation.

In contrast, if the higher-frequency mode, $a_{2}$, is driven, $V(\mathcal{E})$
\emph{always} has a single minimum at $\mathcal{E}=0$ for any values
of system parameters.

The states are not qualitatively changed for finite mechanical and
laser detunings (see Ref.\ \citep{SupplentaryMaterial}).

We now study effects of the dynamically generated synthetic electric
field on light transport. The transmission $T$ is the ratio of the
output power leaking from the non-driven mode, $\kappa|a_{2}|^{2}$
(if mode 1 is driven) or $\kappa|a_{1}|^{2}$ (if mode 2 is driven),
and the driving power $E^{2}/\kappa$. We find that
\begin{equation}
T=\frac{\frac{B^{2}J^{2}}{\kappa^{2}}}{\left(\frac{B^{2}J^{2}}{\kappa^{2}}+\frac{1}{4}\right)^{2}+\frac{1}{4\kappa^{2}}\mathcal{E}^{2}}
\end{equation}
is suppressed when a finite field $\mathcal{E}$ detunes the tunneling
process from resonance. In Figs.~\ref{phdiag}d and ~\ref{phdiag}e,
$\mathcal{E}$ and $T$, respectively, are depicted along the cut
in Fig.~\ref{phdiag}c denoted by the dashed black line.

When light propagates to higher optical frequencies, the phonon-assisted
photon tunneling is suppressed due to the synthetic electric field.
In contrast, the field always vanishes when light propagates towards
lower optical frequencies. In this way, dynamical gauge fields give
rise to a new mechanism for unidirectional light transport. 

\begin{figure}[t]
\centering \includegraphics[width=1\linewidth]{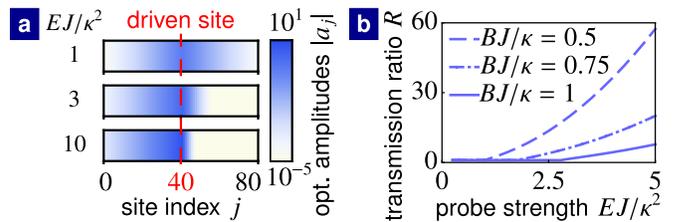} \caption{Light transport in a 1D array with dynamical gauge fields – generation
of a barrier for photon transport induced by synthetic electric fields.
(a) The optical amplitudes $|a_{j}|$ as a function of position for
different values of the laser amplitude $E$ and $BJ/\kappa=1$ (shown
on a logarithmic scale). The dashed red line denotes the driven site.
Transport to the right is strongly suppressed. (b) The ratio $R=|a_{d-1}/a_{d+1}|^{2}$
of the optical amplitudes adjacent to the driven site $j=d$. (Plotted
for $n=81$ sites, site $d=40$ being driven.)}
\label{fig:chain} 
\end{figure}

\emph{Nonlinear unidirectional light transport in a one-dimensional
array}. — The physics of synthetic electric fields also affects photon
transport in arrays (Fig.~\ref{link}c). For more details see Ref.\ \citep{SupplentaryMaterial}.

Fig.~\ref{fig:chain}a shows the result for a 1D array: for a sufficiently
large laser drive, the system switches into a state where finite $\mathcal{E}$
develops to the right of the laser drive. This is the direction where
photons need to gain energy when tunneling, and where we already saw
in the two-site system that (i) a finite field can develop, and (ii)
it suppresses photon transport. In the array, this results in a rapid
exponential suppression of light intensity. In contrast, light easily
propagates towards the left, where $\mathcal{E}$ remains zero. In
Fig.~\ref{fig:chain}b, we plot the ratio $R=|a_{d-1}/a_{d+1}|^{2}$
of transmission to the sites adjacent to the driven site $j=d$ as
a function of $EJ/\kappa^{2}$. The suppression of light propagation
to the right, i.e., $R>1$, is achieved above the threshold of the
laser amplitude. At $m$ sites distance from the driven site, the
ratio is exponentially increased to $R^{m}$.

Our numerical simulations indicate that unidirectional light propagation
can also be observed in two-dimensional square arrays. In the future,
one might study how these phenomena affect synchronization dynamics
of coupled optomechanical self-oscillators \citep{heinrich2011,lauter2015,weiss2017}.

\emph{Experimental parameters required for generating the synthetic
electric field}. — We estimate that unidirectional light transport
can be observed for experimentally realistic parameters. For the membrane-in-the-middle
setup, feasible parameters are $\kappa\approx300\,{\rm kHz}$, $J\approx1\,{\rm Hz}$,
a zero-point fluctuation amplitude of $x_{{\rm ZPF}}\approx10^{-15}\,{\rm m}$
and a number of photons in the cavity $(E/\kappa)^{2}\sim10^{10}$
\citep{sankey2010}. A typical phonon number in limit-cycle oscillations
driven well above threshold is $B^{2}\sim(\kappa/J)^{2}\sim10^{10}$
with a corresponding real oscillation amplitude $2x_{{\rm ZPF}}B\sim100\,{\rm pm}$
\citep{marquardt2006}. Optical modes can be represented by hybridized
modes of a cavity with avoided crossing \citep{sankey2010}. The splitting
of their frequencies $\approx200\:{\rm kHz}$ can match the mechanical
frequency. For these experimental parameters, $EJ/\kappa^{2}\sim1$
and $BJ/\kappa\sim1$ are promising for observing unidirectional light
transport (see Fig.~\ref{phdiag}). The phonon number can be decreased
below the photon number in the driven mode by driving mechanical self-oscillations
closer to threshold \citep{marquardt2006}, fulfilling the necessary
condition for a finite synthetic electric field (Fig.~\ref{phdiag}).

\emph{Conclusions}. — While synthetic gauge fields for photons have
been investigated thoroughly in recent years, little has been known
about the dynamical situation. In this work, we have uncovered how
a synthetic electric field can be spontaneously created in a readily
realizable optomechanical setup. The resulting nonlinear photon-diode
type of unidirectional transport can lead to a large isolation ratio,
especially in arrays. We demonstrate how the interplay of nonlinearity,
dynamics, and artificial gauge fields can produce novel physical effects
and possible new devices.

We thank A. Nunnenkamp and J. Harris for useful comments, and O. Hart
for a careful reading of the manuscript. This work was supported by
the European Union's Horizon 2020 research and innovation programme
under grant agreement No 732894 (FET Proactive HOT).

\bibliographystyle{apsrev4-1}
\addcontentsline{toc}{section}{\refname}\bibliography{references}

\clearpage\onecolumngrid

%%%%%%%%%% Prefix a "S" to all equations, figures, tables and reset the counter %%%%%%%%%% 
\setcounter{equation}{0}
\setcounter{figure}{0}
\setcounter{table}{0}
\setcounter{page}{1}
\makeatletter
\renewcommand{\theequation}{S\arabic{equation}}
\renewcommand{\thefigure}{S\arabic{figure}}
\renewcommand{\bibnumfmt}[1]{[S#1]}
%\renewcommand{\citenumfont}[1]{S#1} 
%%%%%%%%%% Prefix a "S" to all equations, figures, tables and reset the counter %%%%%%%%%%

\begin{center}
\textbf{\large{}Dynamically Generated Synthetic Electric Fields for
Photons – Supplementary material}{\large\par}
\par\end{center}

\section*{Phonon-assisted Photon Tunneling}

In this section, we derive the Hamiltonian (\ref{ham}) that describes
our scenario in the main text including the photon-phonon interaction
term $\hat{b}\hat{a}_{2}^{\dagger}\hat{a}_{1}$. This term is obtained
in optomechanical systems with two (or more) optical modes. Qualitatively,
this term arises whenever there are two optical modes that couple
to the same mechanical resonator, and the term becomes important dynamically
if the mechanical frequency matches the optical frequency difference.
The earliest, well-known example is the membrane-in-the-middle setup
of the Harris group \citep{thompson2008,sankey2010}. More generally,
such a three-body interaction term will generically arise in systems
of nonlinearly coupled modes (e.g. between three optical modes in
a $\chi^{(2)}$-medium, or between microwave modes in the presence
of a Josephson nonlinearity).

We consider a membrane in a cavity whose vibrational mode $\hat{b}$
couples to two optical modes of the cavity. The position of the membrane
determines the frequencies $\omega_{L}$ and $\omega_{R}$ of the
optical modes $\hat{a}_{L}$ (to the left of the membrane) and $\hat{a}_{R}$
(to the right of the membrane), respectively. Placing the membrane
exactly in the middle of the cavity results in equal optical frequencies.
Dislocating the membrane slightly from the center introduces splitting
$\omega=\omega_{R}-\omega_{L}$ between the optical frequencies. Without
loss of generality, we assume $\omega\geq0$. The optical modes $\hat{a}_{L/R}$
coupled to the mechanical mode $\hat{b}$ are described by the Hamiltonian

\begin{equation}
\hat{H}=\omega_{L}\hat{a}_{L}^{\dagger}\hat{a}_{L}+\omega_{R}\hat{a}_{R}^{\dagger}\hat{a}_{R}+\Omega\,\hat{b}^{\dagger}\hat{b}+J_{0}\left(\hat{a}_{L}^{\dagger}\hat{a}_{R}+{\rm h.c.}\right)-g_{0}\left(\hat{a}_{L}^{\dagger}\hat{a}_{L}-\hat{a}_{R}^{\dagger}\hat{a}_{R}\right)\left(\hat{b}^{\dagger}+\hat{b}\right),\label{ham_bare}
\end{equation}
where $\Omega$ is the mechanical frequency, $g_{0}$ is the single-photon
optomechanical coupling strength and $J_{0}$ is the optical coupling
strength. For optical modes with a different transversal spatial profile,
$J_{0}$ can be arbitrarily tuned by the tilt of the membrane \citep{sankey2010}.
Due to the optical coupling, the cavity modes hybridize and that gives
rise to supermodes

\begin{align}
\hat{a}_{1} & =\frac{-J_{0}\thinspace\hat{a}_{L}+\lambda\thinspace\hat{a}_{R}}{\sqrt{J_{0}^{2}+\lambda^{2}}},\label{1 bare}\\
\hat{a}_{2} & =\frac{\lambda\thinspace\hat{a}_{L}+J_{0}\thinspace\hat{a}_{R}}{\sqrt{J_{0}^{2}+\lambda^{2}}},\label{2 bare}
\end{align}
at frequencies $\nu_{1}=\omega_{L}-\lambda$ and $\nu_{2}=\omega_{R}+\lambda$,
where $\lambda=\sqrt{\frac{\omega^{2}}{4}+J_{0}^{2}}-\frac{\omega}{2}$.
In terms of the supermodes, the Hamiltonian reads

\begin{equation}
\hat{H}=\sum_{j=1}^{2}\nu_{j}\hat{a}_{j}^{\dagger}\hat{a}_{j}+\Omega\,\hat{b}^{\dagger}\hat{b}+\left[J_{{\rm res}}\left(\hat{a}_{2}^{\dagger}\hat{a}_{2}-\hat{a}_{1}^{\dagger}\hat{a}_{1}\right)+J\left(\hat{a}_{1}^{\dagger}\hat{a}_{2}+h.c.\right)\right]\left(\hat{b}^{\dagger}+\hat{b}\right),\label{ham_coupled}
\end{equation}
where
\begin{align}
J & =\frac{g_{0}}{\sqrt{1+\frac{\omega^{2}}{4J_{0}^{2}}}},\label{J form}\\
J_{{\rm res}} & =\frac{g_{0}}{\sqrt{1+\frac{4J_{0}^{2}}{\omega^{2}}}}.\label{J res}
\end{align}

Next, we assume driving of $\hat{a}_{L/R}$ with lasers of strengths
$E_{L/R}$ at frequencies $\nu_{D,L/R}$. Neglecting quantum fluctuations
around large optical amplitudes, we can derive equations of motion

\begin{align}
\dot{\phi} & =\Delta_{M}+\frac{J_{{\rm res}}}{B}\,\left(|a_{1}|^{2}-|a_{2}|^{2}\right)\cos\left[\phi-\delta t\right]-\frac{J}{B}{\rm Re}\left[a_{1}^{*}a_{2}e^{-i\phi}+a_{1}a_{2}^{*}e^{-i\phi+2i\delta t}\right],\label{EOM phi full}\\
\dot{a}_{1} & =i\Delta_{1}a_{1}-iE_{1}-iE_{1}^{{\rm res}}e^{-i\delta t}+iB\left[J_{{\rm res}}a_{1}\left(e^{-i\phi+i\delta t}+e^{i\phi-i\delta t}\right)-Ja_{2}\left(e^{-i\phi}+e^{i\phi-i2\delta t}\right)\right]-\frac{\kappa}{2}a_{1},\label{EOM a1 full}\\
\dot{a}_{2} & =i\Delta_{2}a_{2}-iE_{2}+iE_{2}^{{\rm res}}e^{i\delta t}-iB\left[J_{{\rm res}}a_{2}\left(e^{-i\phi+i\delta t}+e^{i\phi-i\delta t}\right)+Ja_{1}\left(e^{i\phi}+e^{-i\phi+2i\delta t}\right)\right]-\frac{\kappa}{2}a_{2},\label{EOM a2 full}
\end{align}
in frames rotating at suitable frequencies (mode $a_{1}$ at $\nu_{D,L}$,
mode $a_{2}$ at $\nu_{D,R}$ and mode $b$ at $\delta=\nu_{D,R}-\nu_{D,L}$),
where we assume mechanical limit-cycle oscillations $\left\langle \hat{b}\right\rangle =Be^{i\phi}$
with a fixed amplitude $B$. We define $\Delta_{1}=\nu_{D,L}-\nu_{1}$,
$\Delta_{2}=\nu_{D,R}-\nu_{2}$, $\Delta_{M}=\delta-\Omega$, $E_{1}=E_{L}/\sqrt{1+\lambda^{2}/J_{0}^{2}}$,
$E_{1}^{{\rm res}}=E_{R}/\sqrt{1+J_{0}^{2}/\lambda^{2}}$, $E_{2}=E_{R}/\sqrt{1+\lambda^{2}/J_{0}^{2}}$
and $E_{2}^{{\rm res}}=E_{L}/\sqrt{1+J_{0}^{2}/\lambda^{2}}$.

Now we consider the resonant case, $\Delta_{1}=\Delta_{2}=0$, when
the generation of the synthetic electric fields is the most pronounced.
In this case, driving of $\hat{a}_{L}$ mostly addresses supermode
$\hat{a}_{1}$, since the laser is on resonance with its frequency
and the overlap of $\hat{a}_{L}$ with the other supermode $\hat{a}_{2}$
is small provided that $\omega\gg J_{0}$. As a result, we can neglect
the residual driving of the supermode $\hat{a}_{2}$. Similarly, driving
of $\hat{a}_{R}$ leads to addressing mostly the supermode $\hat{a}_{2}$.
If the residual drivings are negligible and we tune the mechanical
frequency such that $\Omega=\nu_{2}-\nu_{1}$, the coupling term $\hat{b}\hat{a}_{2}^{\dagger}\hat{a}_{1}$
is selected and the other coupling terms in the Hamiltonian (\ref{ham_coupled})
are off resonance. Neglecting the off-resonant coupling terms within
the rotating-wave approximation, which is valid for $\Omega=\nu_{2}-\nu_{1}\approx\omega\gg\kappa,g_{0},g_{0}B$,
the equations of motion reduce to Eqs.\ (\ref{EOM phi}), (\ref{EOM a1})
and (\ref{EOM a2}) considered in the main text with the effective
tunneling amplitude $J$ given by Eq.\ (\ref{J form}). The tunneling
amplitude $J$ decreases with decreasing $J_{0}.$ As a result, a
fine tuning of the ratio $\omega/J_{0}$ is necessary to achieve the
optimal trade-off between eliminating the residual driving of unwanted
supermodes and maximizing the amplitude of the term $\hat{b}\hat{a}_{2}^{\dagger}\hat{a}_{1}$.

We have derived the Hamiltonian (\ref{ham}) for optical supermodes,
which is considered as the starting point in the main text, from the
fundamental optomechanical Hamiltonian (\ref{ham_bare}) for two optical
modes coupled to a single mechanical mode. In summary, whenever two
optical modes and one mechanical resonator are in a mutual interaction,
and when the mechanical frequency matches the optical frequency difference
(at least approximately), the interaction term $\hat{a}_{2}^{\dagger}\hat{a}_{1}\hat{b}+{\rm h.c.}$
assumed in our work is the generic outcome.

\section*{Steady states of the two-site system}

\label{steady states} In this section, we analyze the steady states
of the two-site system with single mode being driven. They are stationary
solutions of the equations of motion (Eqs.\ (\ref{EOM phi}), (\ref{EOM a1})
and (\ref{EOM a2}) in the main text) constant in time. We first apply
a time-dependent gauge transformation to express the time-evolution
of the mechanical phase in a form of an effective optical frequency
shift. Then we find a stationary condition for the synthetic electric
field $\mathcal{E}$. Finally, we use an effective potential for the
synthetic electric field to study stability of its stationary solutions.

As mentioned in the main text, we assume that only one mode is driven.
We label the driven mode by the index $k=1,2$. Driving strengths
can then be expressed as $E_{j}=E\,\delta_{j,k}$ for $j=1,2$, where
$\delta_{j,k}$ is the Kronecker delta. The detuning of the non-driven
mode can be set to zero, since there is no driving frequency. Therefore,
the optical detunings can be expressed as $\Delta_{j}=\Delta_{{\rm O}}\,\delta_{j,k}$.
We make use of the time-dependent gauge transformation
\begin{align}
\phi & =\tilde{\phi}+\chi,\label{gauge A phi}\\
a_{1} & =\tilde{a}_{1}e^{-i\chi\,\delta_{2,k}},\label{gauge A a1}\\
a_{2} & =\tilde{a}_{2}e^{i\mathcal{\chi}\,\delta_{1,k}},\label{gauge A a2}
\end{align}
which moves the dynamics of the mechanical phase to the time-dependent
gauge parameter $\chi$. By appropriately choosing $\chi,$ we can
always achieve $\tilde{\phi}={\rm 0}$. The time-dependent gauge transformation
leaves the absolute values of the optical amplitudes unchanged. As
a result, a particular value of the gauge parameter is irrelevant.
Only its first derivative $\dot{\chi}=\dot{\phi}$ influences the
optical occupations. The time evolution of the mechanical phase results
in an effective shift $\left(\delta_{2,k}-\delta_{1,k}\right)\mathcal{\dot{\chi}}$
of the non-driven optical mode's frequency. Note that the driven mode,
$a_{k}$, is forced to oscillate with the frequency of the laser drive,
and thus it does not experience any frequency shift.

The role of the optical frequency shift $\dot{\chi}$ can be understood
in analogy to electromagnetism. The mechanical phase corresponds to
an effective vector potential. According to conventional electromagnetism,
the time evolution of the vector potential generates an electric field.
This electric field can be also represented by a scalar potential
gradient. In this analogy, the time evolution of the mechanical phase
generates a synthetic electric field $\mathcal{E}=\dot{\chi}$ for
photons, which represent an effective optical frequency shift.

To provide the fixed point analysis for the both cases $k=1,2$ at
once, we use general indexes $(k,l)\in\left\{ (1,2),(2,1)\right\} $
to label the optical modes. According to the gauge transformation
(\ref{gauge A phi}), (\ref{gauge A a1}), and (\ref{gauge A a2}),
the equations of motion transform to 
\begin{align}
\dot{\tilde{\phi}} & =\mathcal{E}-\Delta_{{\rm M}}+\frac{J}{B}\,{\rm Re}\left[\tilde{a}_{k}^{*}\tilde{a}_{l}\right]=0,\label{phi tilde}\\
\dot{\tilde{a}}_{k} & =i\Delta_{{\rm O}}\tilde{a}_{k}-iE-iJB\tilde{a}_{l}-\frac{\kappa}{2}\tilde{a}_{k},\label{a1 tilde}\\
\dot{\tilde{a}}_{l} & =i\left(\delta_{2,k}-\delta_{1,k}\right)\mathcal{E}\tilde{a}_{l}-iJB\tilde{a}_{k}-\frac{\kappa}{2}\tilde{a}_{l},\label{a2 tilde}
\end{align}
where we substituted $\mathcal{E}=\dot{\chi}$ . Taking the time derivative
of Eq.~(\ref{phi tilde}), we obtain the equation of motion for the
synthetic electric field 
\begin{equation}
\dot{\mathcal{E}}=-\kappa\left(\mathcal{E}-\Delta_{{\rm M}}\right)+\frac{EJ}{B}\,{\rm Im}\left[\tilde{a}_{l}\right]+\frac{J}{B}\left(\left(\delta_{2,k}-\delta_{1,k}\right)\mathcal{E}-\Delta_{{\rm O}}\right){\rm Im}\left[\tilde{a}_{k}^{*}\tilde{a}_{l}\right].\label{el}
\end{equation}
To find stationary solutions of the equations of motion (\ref{a1 tilde}),
(\ref{a2 tilde}), and (\ref{el}), we first use that the equations
(\ref{a1 tilde}) and (\ref{a2 tilde}) are linear in terms of optical
amplitudes. For a given value of the synthetic electric field $\mathcal{E}$,
the stationary optical amplitudes read 
\begin{align}
\tilde{a}_{k} & =E\frac{\left(\delta_{2,k}-\delta_{1,k}\right)\mathcal{E}+i\frac{\kappa}{2}}{-J^{2}B^{2}+\left(\delta_{2,k}-\delta_{1,k}\right)\mathcal{E}\Delta_{{\rm O}}-\left(\frac{\kappa}{2}\right)^{2}+i\frac{\kappa}{2}\left[\left(\delta_{2,k}-\delta_{1,k}\right)\mathcal{E}+\Delta_{{\rm O}}\right]},\label{a1 stat}\\
\tilde{a}_{l} & =\frac{JB}{\left(\delta_{2,k}-\delta_{1,k}\right)\mathcal{E}+i\frac{\kappa}{2}}\tilde{a}_{k}.\label{a2 stat}
\end{align}

\begin{figure}[t]
\centering \includegraphics[width=0.5\linewidth]{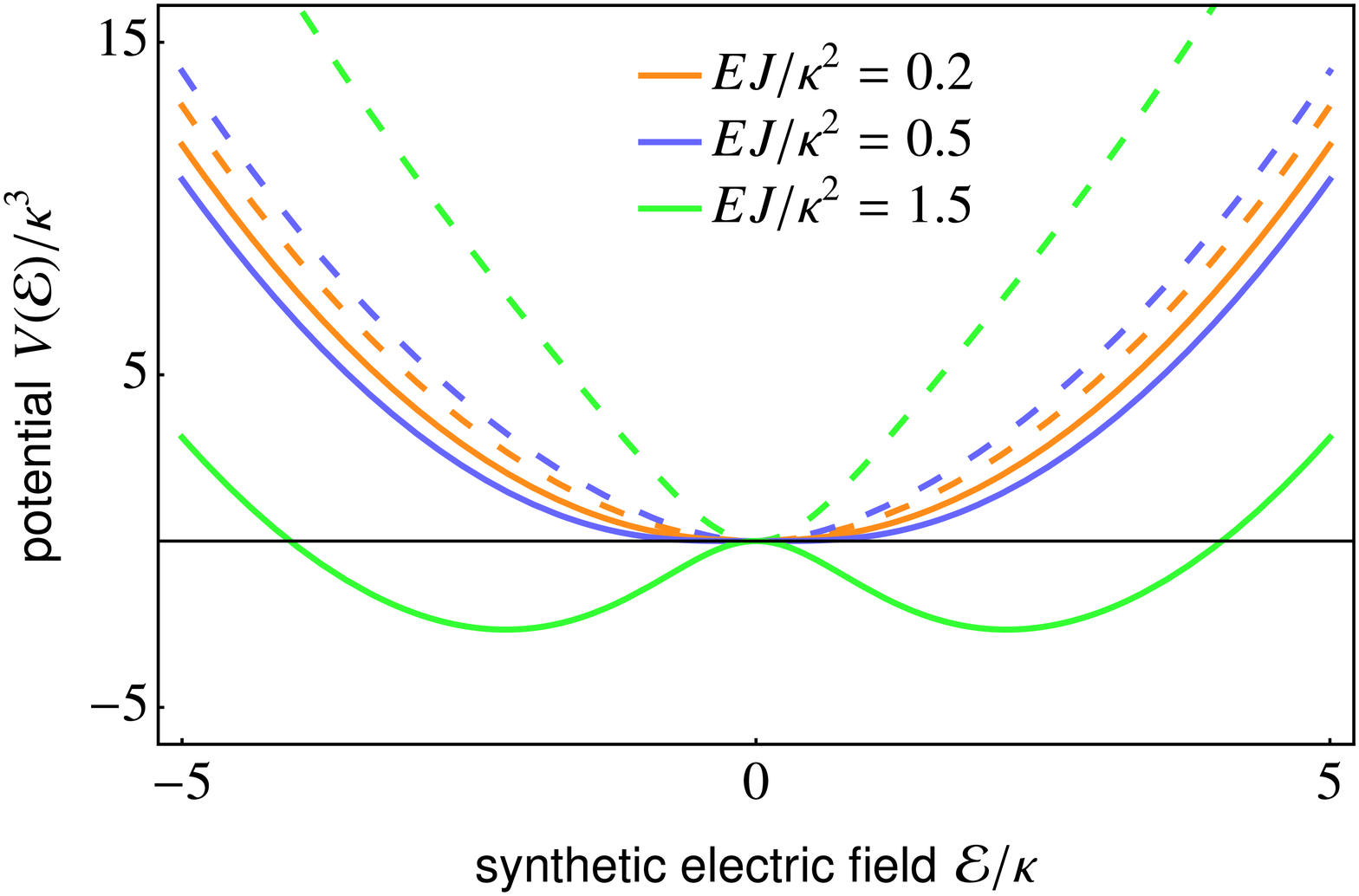} \caption{The potential for the synthetic electric field. It has a single minimum
at $\mathcal{E}=0$ for $k=2$ (dashed lines) when the higher optical
frequency is driven. For $k=1$, when the lower optical frequency
is driven, the stationary value $\mathcal{E}=0$ becomes unstable
with increasing $EJ/\kappa^{2}$ as two minima with a finite frequency
shift emerge (solid lines). (The potential is plotted for $BJ/\kappa=0.5$.)}
\label{potential} 
\end{figure}

In following, we set $\Delta_{{\rm O}}=\Delta_{{\rm M}}=0$ to present
the important features of the steady states. These features are not
changed by finite detunings. We discuss the effects of finite detunings
at the end of this section. By substituting the stationary values
of the optical amplitudes (\ref{a1 stat}), (\ref{a2 stat}) into
Eq.~(\ref{el}), we obtain the stationary condition for the synthetic
electric field 
\begin{equation}
0=\dot{\mathcal{E}}=-\kappa\mathcal{E}\,\frac{\left(\frac{\mathcal{E}}{\kappa}\right)^{2}+4\left[\left(\frac{JB}{\kappa}\right)^{2}+\frac{1}{4}\right]^{2}+4\left(\delta_{2,k}-\delta_{1,k}\right)\left(\frac{EJ}{\kappa^{2}}\right)^{2}}{\left(\frac{\mathcal{E}}{\kappa}\right)^{2}+4\left[\left(\frac{JB}{\kappa}\right)^{2}+\frac{1}{4}\right]^{2}}.\label{stat el}
\end{equation}
For $k=2$, when the mode with the higher optical frequency, $a_{2}$,
is driven, only the single stationary solution, $\mathcal{E}=0$,
exists. For $k=1$, when the mode with the lower optical frequency,
$a_{1}$, is driven, stationary solutions with a finite synthetic
electric field 
\begin{equation}
\mathcal{E}_{\pm}=\pm2\kappa\sqrt{\left(\frac{EJ}{\kappa^{2}}\right)^{2}-\left[\left(\frac{JB}{\kappa}\right)^{2}+\frac{1}{4}\right]^{2}}
\end{equation}
emerge, in addition to $\mathcal{E}=0$, for $4EJ/\kappa^{2}>4\left(BJ/\kappa\right)^{2}+1$.

%potential
To gain intuition about the stability of these stationary solutions,
we find the potential 
\begin{equation}
V(\mathcal{E})=\frac{\kappa^{3}}{2}\left(\left(\frac{\mathcal{E}}{\kappa}\right)^{2}+4\left(\delta_{2,k}-\delta_{1,k}\right)\left(\frac{JE}{\kappa^{2}}\right)^{2}\ln\left(\left(\frac{\mathcal{E}}{\kappa}\right)^{2}+4\left[\left(\frac{JB}{\kappa}\right)^{2}+\frac{1}{4}\right]^{2}\right)\right),
\end{equation}
such that $-{\rm d}V(\mathcal{E})/{\rm d}\mathcal{E}$ is equal to
the right hand side of Eq.~(\ref{stat el}). The potential shows
that the stationary solution $\mathcal{E}=0$ is always a stable steady
state for $k=2$ when the optical mode with the higher optical frequency
is driven (see Fig.~\ref{potential}). The stability of the steady
state does not depend on the system parameters. For $k=1$, when the
mode with the lower optical frequency is driven, the stability of
the steady state depends on the two dimensionless parameters $EJ/\kappa^{2}$
and $BJ/\kappa$. The potential in Fig.~\ref{potential} shows that
the steady state $\mathcal{E}=0$ is the only stationary solution
and it is stable in the white region of the phase diagram depicted
in Fig.~\ref{phdiag} of the main text. It becomes unstable as the
two steady states with a finite synthetic electric field emerge in
the colored region of the phase diagram in Fig.~\ref{phdiag} of
the main text. Note that the potential does not provide conclusive
information about the stability of the steady states because it does
not take into account the dynamics of the optical modes. Therefore,
the linear stability analysis was used to confirm that the stability
of the steady states is determined correctly by the potential $V(\mathcal{E})$.

A finite mechanical detuning, $\Delta_{{\rm M}}\neq0$, detunes the
phonon-assisted photon tunneling process from resonance. In this way,
the mechanical detuning represents a static synthetic electric field
for photons in contrast to the dynamically generated synthetic electric
field $\mathcal{E}$. If the higher optical frequency is driven, the
dynamically generated synthetic electric field $\mathcal{E}$ acts
against this static synthetic electric field and increases transmission
to the lower optical frequency with the increasing laser amplitude.
On the other hand, for the lower optical frequency being driven, the
dynamically generated synthetic electric field detunes the tunneling
process further from resonance with the increasing laser amplitude.
As a result, it decreases light propagation to the non-driven optical
mode. Above some threshold of the laser amplitude, the synthetic electric
field bifurcates as the effective potential have two local minima.
This again happens only for the lower frequency being driven.

A finite laser detuning, $\Delta_{{\rm O}}\neq0$, suppresses the
coherent driving, which results in a smaller optical amplitude of
the driven mode. For the higher optical frequency being driven, the
synthetic electric field always vanishes even for a finite optical
detuning. It vanishes also when the lower optical frequency is driven
for small laser amplitudes. Similarly as in the resonant case, the
synthetic electric field bifurcates to finite values over the threshold
of the laser amplitude for the lower frequency being driven. The threshold
and the values of the bifurcated synthetic electric field are modified
by the finite optical detuning since it changes the population and
the phase of the driven optical mode. However, the qualitative features
of the synthetic electric field remain the same. The synthetic electric
field is generated only above threshold and only for the lower optical
frequency being driven.

\section*{Numerical simulations of the full equations of motion}

In this section, we present numerical simulations of the system described
by the fundamental Hamiltonian (\ref{ham_bare}) when we consider
the laser drive coupling to the original (uncoupled) cavity modes,
out of which the supermodes are formed. In this way, we demonstrate
the validity of the results presented in the main text where the description
of the system is simplified by assuming that individual supermodes
can be separately coupled to the laser drive.

\begin{figure}[t]
\centering \includegraphics[width=0.5\linewidth]{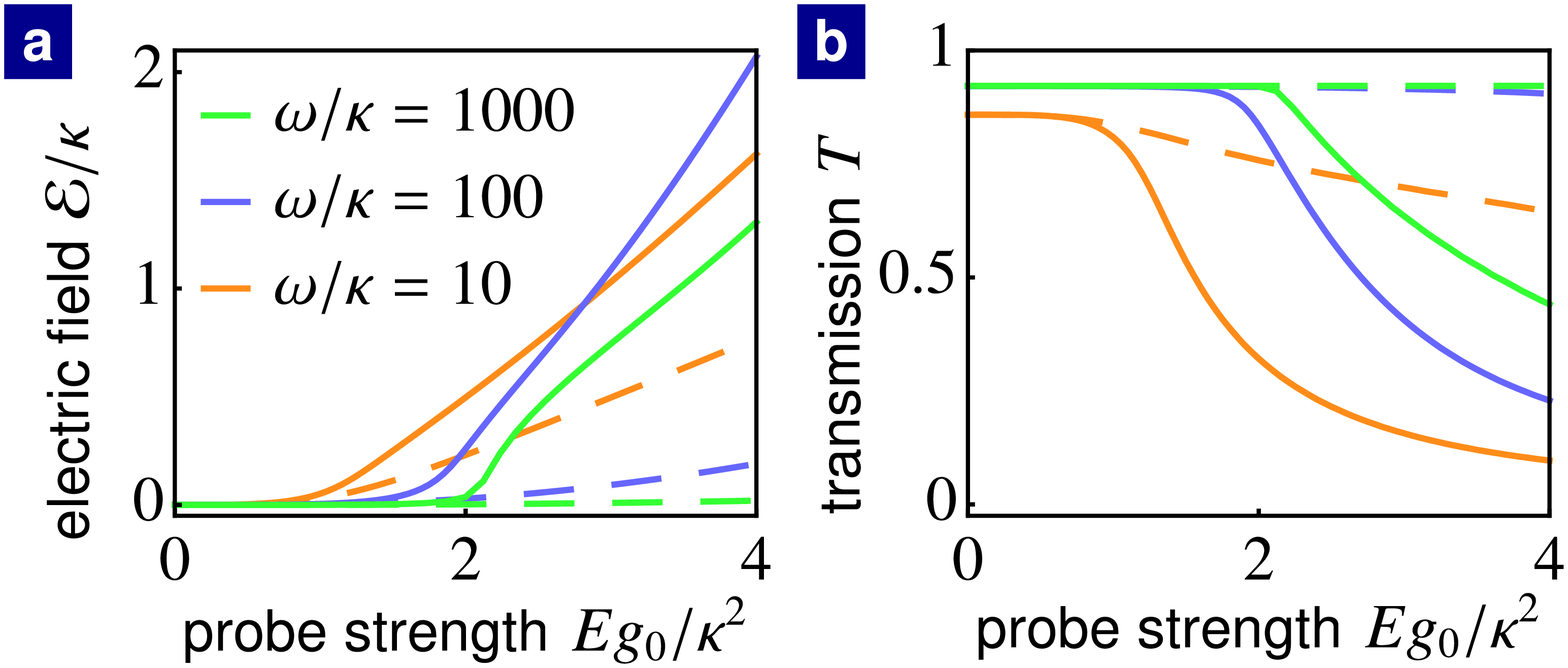}
\caption{Dynamically generated synthetic electric fields in
the two site system considering the model described by the fundamental
Hamiltonian (\ref{ham_bare}). (a) For the uncoupled optical mode,
$a_{L}$, being driven with a laser at frequency $\nu_{1}$ (solid
lines), a large synthetic electric field $\mathcal{E}$ develops.
For driving the uncoupled optical mode, $a_{R}$, with a laser at
frequency $\nu_{2}$ (dashed lines), a small synthetic electric field
develops, however, it is reduced as the sideband ratio $\omega/\kappa$
increases (see the color scale). (b) As a result, the optical transmission
$T$ to the right (solid lines) is significantly suppressed in comparison
to the transmission to the left (dashed lines). In a one-dimensional
array, this transmission ratio gets exponentiated by the length of
the array, leading to a very significant suppression of transport
in one direction. (Plotted for $Bg_{0}/\kappa=2,$ $J_{0}/\omega=0.085$.)}
\label{bare modes}
\end{figure}

We now simulate the dynamics of the uncoupled optical modes $\hat{a}_{L}$
and $\hat{a}_{R}$ according to the fundamental Hamiltonian (\ref{ham_bare}).
To this end, we derive classical equations of motion

\begin{align}
\dot{\phi} & =-\Omega+\frac{g_{0}}{B}\,\left(|a_{L}|^{2}-|a_{R}|^{2}\right)\cos\phi,\label{EOM phi-1}\\
\dot{a}_{L} & =i\left(\Delta_{L}+2g_{0}B\cos\phi\right)a_{L}-iE_{L}-iJ_{0}a_{R}-\frac{\kappa}{2}a_{L},\label{EOM a1-1}\\
\dot{a}_{R} & =i\left(\Delta_{R}-2g_{0}B\cos\phi\right)a_{R}-iE_{R}-iJ_{0}a_{L}-\frac{\kappa}{2}a_{R},\label{EOM a2-1}
\end{align}
neglecting quantum fluctuations around the expectation values $a_{L/R}=\langle\hat{a}_{L/R}\rangle$,
where we again assume that the mechanical mode, $\left\langle \hat{b}\right\rangle =Be^{i\phi}$,
performs limit-cycle oscillation with a fixed amplitude $B$ as well
as $\Delta_{L}=\nu_{D}-\omega_{L}$ , $\Delta_{R}=\nu_{D}-\omega_{R}$,
and $\nu_{D}$ is the frequency of the laser drive. We consider driving
of a single uncoupled optical mode at resonance with the corresponding
supermode,\ i.e. $\nu_{D}=\nu_{1}$ for $E_{L}\neq0$ or $\nu_{D}=\nu_{2}$
for $E_{R}\neq0$, and $\Omega=\nu_{2}-\nu_{1}.$ The generated synthetic
electric field and the optical transmission are shown in Fig.\ \ref{bare modes}
as a function of the rescaled driving strength $Eg_{0}/\kappa^{2}$.
One can see in Fig.\ \ref{bare modes}a that a large synthetic electric
field is generated for mode $a_{L}$ being driven (solid lines). As
a result, the transmission to the right (solid lines) is significantly
suppressed, see Fig.\ \ref{bare modes}b.

For driving mode $a_{R}$, a small synthetic electric field (dashed
lines) is generated. This is in contrast to the simplified model in
the main text, where the synthetic electric field completely vanishes
when light propagates to the lower optical frequency. The small generated
synthetic electric field is a result of the residual driving of the
supermode $a_{1}$ due to its non-vanishing overlap with the driven
uncoupled mode $a_{R}$. Since the mechanical frequency, $\Omega=\nu_{2}-\nu_{1}$,
is chosen to match the optical frequency difference, the supermode
$a_{1}$ is driven on the blue sideband. However, the residual driving
can be suppressed by increasing the sideband ratio $\omega/\kappa$,
see Fig.\ \ref{bare modes}a. As a result, a significant suppression
of the optical transmission to the right (solid lines) in comparison
to the transmission to the left (dashed lines) can be reached, see
Fig.\ \ref{bare modes}b. This leads to unidirectional transport
of light which works especially well in one-dimensional arrays. In
such an array, the transmission ratio is exponentiated by the length
of the array, which results in a large suppression of transport in
one direction.

Simulating the dynamics of the uncoupled optical modes, we have shown
that unidirectional light transport via synthetic electric fields
is achieved for the fundamental model described by the Hamiltonian
(\ref{ham_bare}). This demonstrates that the model in the main text
indeed captures the important features of the interaction between
the two optical modes and the mechanical mode in our scenario. Our
results show that unidirectional light transport is more pronounced
with the increasing sideband ratio $\omega/\kappa$.

\section*{Phase locking}

In this section, we provide a brief summary of phase locking, which
can be reached in the two site system by simultaneously driving both
optical modes. Note that the analysis presented in the main text is
for a single mode driven only. We present here quantitative features
of phase locking, which has been previously well studied in a similar
optomechanical system \citep{heinrich2011}.

The starting point of the analysis are the equations of motion (\ref{EOM phi}),
(\ref{EOM a1}) and (\ref{EOM a2}) in the main text. The stationary
values for the optical amplitudes are 

\begin{align}
a_{1} & =-\frac{JBE_{2}e^{-i\phi}+i\frac{\kappa}{2}E_{1}}{J^{2}B^{2}+\left(\frac{\kappa}{2}\right)^{2}},\label{a1 stat lock}\\
a_{2} & =-\frac{JBE_{1}e^{i\phi}+i\frac{\kappa}{2}E_{2}}{J^{2}B^{2}+\left(\frac{\kappa}{2}\right)^{2}}.\label{a2 stat lock}
\end{align}
Note that if both optical modes are driven, the phases $\varphi_{1}$
and $\varphi_{2}$ of the laser amplitudes $E_{1}$ and $E_{2}$,
respectively, determine the phases $\theta_{1}$ and $\theta_{2}$
of the intracavity modes. This is different to the case when only
a single optical mode is driven, where the phase of the driving amplitude
is irrelevant. 

The stationary value of the mechanical phase $\phi$ obeys the Adler
equation

\begin{equation}
\Delta_{M}-|a_{1}||a_{2}|\cos(\phi-\theta)=0.\label{adler-1}
\end{equation}
where $\theta=\theta_{2}-\theta_{1}.$ However, the absolute values
of the optical amplitudes $|a_{1}|$ and $|a_{2}|$ depend on the
phase difference $\phi-\theta$. The Adler equation still determines
uniquely the stationary value of $\cos(\phi-\theta)$ but the full
analytical expression of this equation is complicated. Thus it is
simpler to switch to the phase difference $\varphi=\varphi_{2}-\varphi_{1}$
of the laser phases $\varphi_{1}$ and $\varphi_{2}$. The Adler equation
then has the form

\begin{equation}
\Delta_{M}-\frac{J}{B}\frac{|E_{1}||E_{2}|}{J^{2}B^{2}+\left(\frac{\kappa}{2}\right)^{2}}\cos(\phi-\varphi)=0.\label{adler}
\end{equation}
We can easily read off that the stationary solution of $\phi$ exists
for 

\begin{equation}
|\Delta_{M}|\leq\frac{J}{B}\frac{|E_{1}||E_{2}|}{J^{2}B^{2}+\left(\frac{\kappa}{2}\right)^{2}}.\label{adler2}
\end{equation}
The mechanical phase $\phi$ is locked under this condition to the
difference $\varphi$ of the laser drives' phases. Since there is
one-to-one correspondence between the laser drives' phases and the
intracavity modes' phases, the mechanical phase $\phi$ can be equivalently
though to be locked to the phase difference $\theta$ of the intracavity
modes.

\section*{one-dimensional arrays}

\label{chain} Here we provide details about one-dimensional arrays
analyzed in the main text. We consider an array, depicted in Fig.~\ref{link}c
of the main text, represented by a stack of membranes inside a cavity.
The sites of the array support optical modes $a_{j}$ whose frequencies
$\nu_{j}$ increase with site index $j=1,..,n$. We assume that the
phonon-assisted photon tunneling processes are resonant: $\Omega_{j}=\nu_{j+1}-\nu_{j}$,
where $\Omega_{j}$ is the frequency of the mechanical oscillator
assisting tunneling between modes $\hat{a}_{j}$ and $\hat{a}_{j+1}$.
Specifically, we will consider a situation where some optical mode
$j=d$ is driven resonantly from the side, to study light propagation
towards the left ($j<d$), and towards the right ($j>d$). Alternatively
to membrane stacks, suitably designed coupled cavity arrays in optomechanical
crystals could implement such a setup.

The mechanical oscillators are again assumed to perform limit cycle
oscillations $\langle\hat{b}_{j}\rangle=B\,e^{i\phi_{j}}$ with free
phases and with a fixed amplitude $B$ equal for all mechanical oscillators.
By straightforward extension of Eqs.~(\ref{EOM phi}), (\ref{EOM a1}),
and (\ref{EOM a2}) (in the main text), we obtain the coupled equations
of motion for the optical amplitudes and the mechanical phases 
\begin{align}
\dot{\phi}_{j} & =-\frac{J}{B}\,{\rm Re}\left[a_{j}^{*}a_{j+1}e^{-i\phi_{j}}\right],\label{array phase}\\
\dot{a}_{j} & =-iE_{j}\delta_{j,d}-iJBe^{-i\phi_{j}}a_{j+1}-iJBe^{i\phi_{j-1}}a_{j-1}-\frac{\kappa}{2}a_{j},\label{array a}
\end{align}
where $\delta_{j,d}$ is the Kronecker delta. The optical modes are
expressed in the frames rotating with their frequencies $\nu_{j}$
and the mechanical modes are in the frames rotating with the difference
of optical frequencies on the neighboring sites: $\nu_{j+1}-\nu_{j}$.

We study the dynamics of one-dimensional arrays by numerically solving
the classical equations of motion (\ref{array phase}) and (\ref{array a}).
The system converges to a steady state for any values of the parameters
$EJ/\kappa^{2}$ and $BJ/\kappa$. Properties of the steady states
are discussed in the main text.
\end{document}